\documentclass[12pt]{article}
\usepackage{oldgerm}
\usepackage{yfonts}
\usepackage{euler}
\usepackage{graphics}
\usepackage{bm}
\usepackage{graphicx}
\usepackage{amsmath}
\usepackage{amssymb}
\usepackage{amstext}
\usepackage{amscd}
\usepackage{amsfonts}
\usepackage{appendix}
\usepackage{wasysym}
\usepackage{epstopdf}
\usepackage{color}
\DeclareMathAlphabet{\EuFrak}{U}{euf}{m}{n}
\DeclareMathAlphabet{\EuScript}{U}{eus}{m}{n}

\newcommand{\nd}{\noindent}

\newcommand{\be}{\begin{equation}}
\newcommand{\ee}{\end{equation}}
\newcommand{\ben}{\begin{eqnarray}}
\newcommand{\een}{\end{eqnarray}}

\title{{\bf On the putative essential discreteness of
q-generalized entropies}}
\author{{A. Plastino$^1$ and M. C. Rocca$^1$} \\
\small{$^1$ La Plata National University and
Argentina's National Research Council}\\
\small{(IFLP-CCT-CONICET)-C. C. 727, 1900 La Plata - Argentina}}

\date{\today}

\begin{document}

\maketitle

\begin{abstract}

\nd \small{It has been argued in  [EPL {\bf 90} (2010)  50004], entitled {\it Essential discreteness in generalized thermostatistics with non-logarithmic entropy}, that
''continuous Hamiltonian systems with long-range interactions and the so-called q-Gaussian momentum
distributions are seen to be outside the scope of non-extensive statistical mechanics''. The arguments are clever and appealing. We show here that, however, some mathematical subtleties render them unconvincing
}

\nd Keywords: MaxEnt, functional variation, measures, q-statistics.

\end{abstract}

\newpage

\renewcommand{\theequation}{\arabic{section}.\arabic{equation}}

\section{Introduction}

\setcounter{equation}{0}

\nd During more than 25 years, an important topic in statistical
mechanics theory revolved around the notion of generalized
q-statistics, pioneered by Tsallis \cite{tsallis88}. It has been
amply demonstrated that, in many occasions,  the celebrated
Boltzmann-Gibbs logarithmic entropy does not yield a correct
description of the system under scrutiny \cite{tsallisbook}. Other
entropic forms, called q-entropies, produce a much better
performance \cite{tsallisbook}. One may cite a large number of
such instances. For example, non-ergodic systems exhibiting  a
complex dynamics \cite{tsallisbook}.

\nd The non-extensive statistical mechanics of Tsallis' has been
employed to fruitfully discuss phenomena in variegated fields. One
may mention, for instance, high-energy physics
\cite{[4]}-\cite{[44]}, spin-glasses \cite{[5]}, cold atoms in
optical lattices \cite{[6]}, trapped ions \cite{[7]}, anomalous
diffusion \cite{[8]}, \cite{[9]}, dusty plasmas \cite{[10]},
low-dimensional dissipative and conservative maps in  dynamical
systems \cite{[11]}, \cite{[12]}, \cite{[13]}, turbulent flows
\cite{[14]}, Levy flights \cite{wilk}, the QCD-based Nambu, Jona,
Lasinio  model of a many-body field theory \cite{15}, etc. Notions
related to q-statistical mechanics have been found useful not only
in physics but also in chemistry, biology, mathematics, economics,
and informatics \cite{[17]}, \cite{[18]}, \cite{[19]}.

\vskip 3mm
 \nd In this work we revisit results presented in \cite{abe}. First, we note that \cite{abe} has been criticized, in a manner unrelated to ours here, in a Comment \cite{Andresen}. There is also a reply by Abe to that Comment \cite{aberep}.  
 It is stated in  \cite{abe} that
one encounters an essential discreteness in generalized thermostatistics with non-logarithmic entropy. Thus,
 ''continuous Hamiltonian systems with long-range interactions and the so-called q-Gaussian momentum
distributions are seen to be outside the scope of non-extensive statistical mechanics'' \cite{abe}. The pertinent arguments are clever and appealing. However, as we will show here that,  some mathematical subtleties render them unconvincing. The main reason is the Functional Analysis  is the branch of mathematics operative in this context, not simple Calculus. Functional variational procedures
are described, for instance, in Ref. \cite{shilov}.
\setcounter{equation}{0}

\section{Continuous variational Tsallis' case}

\nd The functional MaxEnt treatment is given in \cite{2ndvariation}.
 Let $P$ stand for the pertinent probability distribution (PD). One evaluates mean values here  in the customary fashion, linear in
$P$, i.e., \newline $<R>= \int_M R P \, d\mu$. It is well known that the MaxEnt variational Tsallis functional
 is \cite{tsallisbook}
\begin{equation}
\label{eq2.1} F_S(P)=-\int\limits_M\,
P^q\ln_q(P)\;d\mu+\alpha\left(
\int\limits_MPH\;d\mu-<U>\right)+\gamma
\left(\int\limits_MP\;d\mu-1\right).
\end{equation}
For the variational increment $h$  we have  \cite{shilov}
\[F_S(P+h)-F_S(P)=-\int\limits_M(P+h)^q\ln_q(P+h)\;
d\mu+\alpha\left[
\int\limits_M(P+h)H\;d\mu-<U>\right]+\]
\[\gamma\left[\int\limits_M(P+h)\;d\mu-1\right]+
\int\limits_MP^q\ln_q(P)\;d\mu-\alpha\left(
\int\limits_MPH\;d\mu-<U>\right)-\],
\begin{equation}
\label{eq2.2}
\gamma\left(\int\limits_MP\;d\mu-1\right).
\end{equation}
Eq.  (\ref{eq2.2}) can be recast as
\[F_S(P+h)-F_S(P)=
\int\limits_M\left[\left(\frac {q} {1-q}\right)P^{q-1}
+\alpha H +\gamma\right]h\;d\mu-\]
\begin{equation}
\label{eq2.3} \int\limits_MqP^{q-2}\frac {h^2}
{2}\;d\mu+O(h^3).
\end{equation}
Eq. (\ref{eq2.3}) leads now to the following equations
\begin{equation}
\label{eq2.4} \left(\frac {q} {1-q}\right)P^{q-1} +\alpha H
+\gamma=0,
\end{equation}
\begin{equation}
\label{eq2.5} -\int\limits_MqP^{q-2}h^2\;d\mu\leq C||h||^2.
\end{equation}
Eq. (\ref{eq2.4}) is the  Euler-Lagrange one while (\ref{eq2.5})
gives bounds originating from the second variation  \cite{shilov}. Starting with (\ref{eq2.4}) we use the procedure given in \cite{tsallisbook}. One first gives the Lagrange multipliers $\alpha$ and $\beta$ a prescribed {\it form} in terms of a (thus far unknown) quantity $Z$: 

\begin{equation}
\label{eq2.6} \alpha=\beta q  Z^{1-q},
\end{equation}
\begin{equation}
\label{eq2.7}
\gamma=\frac {q} {q-1}Z^{1-q},
\end{equation}
and then $Z$ is determined  by appeal to normalization. Accordingly, one has
\begin{equation}
\label{eq2.8}
P=\frac {[1+\beta(1-q)H]^{\frac {1} {q-1}}} {Z} =
e_{2-q}(-\beta H)/Z,
\end{equation}
and, on account of normalization, 
\begin{equation}
\label{eq2.9}
Z=\int\limits_M [1+\beta(1-q)H]^{\frac {1}
{q-1}}\;d\mu.
\end{equation}
 For Eq. (\ref{eq2.9}) we have,
\be
\label{eq2.10}
W=-\int\limits_MqP^{q-2}h^2\;d\mu=
-\int\limits_MqZ^{2-q} [1+\beta(1-q)H]^{\frac {q-2} {q-1}}
h^2\;d\mu.
\ee
How  to obtain a bound is discussed in \cite{2ndvariation}.

\setcounter{equation}{0}

\section{Discrete variational Tsallis' case}

The concomitant
Tsallis discrete  functional is
\begin{equation}
\label{eq3.1}
F_S(P)=-\sum\limits_{i=1}^n
P_i^q\ln_q(P_i)+\lambda_1\left(
\sum\limits_{i=1}^{n}P_iU_i-<U>\right)+\lambda_2
\left(\sum\limits_{i=1}^{n}P_i-1\right)
\end{equation}
For the increment we have
\[F_S(P+h)-F_S(P)=-\sum\limits_{i=1}^n(P_i+h_i)^q\ln_q(P_i+h_i)
+\lambda_1\left[
\sum\limits_{i=1}^n(P_i+h_i)U_i-<U>\right]+\]
\[\lambda_2\left[\sum\limits_{i=1}^n(P_i+h_i)-1\right]+
\sum\limits_{i=1}^nP_i^q\ln_q(P_i)-\lambda_1\left(
\sum\limits_{i=1}^nP_iU_i-<U>\right)-\]
\begin{equation}
\label{eq3.2}
\lambda_2\left(\sum\limits_{i=1}^nP_i-1\right)
\end{equation}
Eq.  (\ref{eq3.2}) can be recast as
\[F_S(P+h)-F_S(P)=
\sum\limits_{i=1}^n\left[\left(\frac {q} {1-q}\right)P_i^{q-1}
+\lambda_1 U_i +\lambda_2\right]h_i\]
\begin{equation}
\label{eq3.3} -\sum\limits_{i=1}^nqP_i^{q-2}\frac {h_i^2}
{2}+O(h^3),
\end{equation}
Eq. (\ref{eq3.3}) leads to the following equations:
\begin{equation}
\label{eq3.4} \left(\frac {q} {1-q}\right)P_i^{q-1} +\lambda_1 U_i
+\lambda_2=0,
\end{equation}
\begin{equation}
\label{eq3.5} -\sum\limits_{i=1}^nqP_i^{q-2}h_i^2\leq C||h||^2
\end{equation}
Eq. (\ref{eq3.4}) is the  Euler-Lagrange one while (\ref{eq3.5})
gives bounds originating from the second variation.

\nd  Starting with (\ref{eq3.4}), we use again the procedure given in \cite{tsallisbook} and in Section 2. One first gives the Lagrange multipliers $\lambda_1$ and $\lambda_2$ a prescribed {\it form} in terms of a (thus far unknown) quantity $Z$: 
\begin{equation}
\label{eq3.6} \lambda_1=\beta q  Z^{1-q},
\end{equation}
\begin{equation}
\label{eq3.7}
\lambda_2=\frac {q} {q-1}Z^{1-q},
\end{equation}
and then has
\begin{equation}
\label{eq3.8}
P_i=\frac {[1+\beta(1-q)U_i]^{\frac {1} {q-1}}} {Z},
\end{equation}
so that normalization demands that
\begin{equation}
\label{eq3.9}
Z=\sum\limits_{i=1}^n [1+\beta(1-q)U_i]^{\frac {1}
{q-1}}.
\end{equation}
For Eq. (\ref{eq3.5}) we have,
\begin{equation}
\label{eq3.10}
W=-\sum\limits_{i=1}^nqP_i^{q-2}h_i^2=
-\sum\limits_{i=1}^nqZ^{2-q} [1+\beta(1-q)U_i]^{\frac {q-2} {q-1}}
h_i^2.
\end{equation}
Just how to obtain a bound follows the lines discussed in \cite{2ndvariation}, adapted
to the discrete scenario.

\setcounter{equation}{0}

\section{Our above results vis-a-vis those of Ref. \cite{abe}}

We have seen in the two previous Sections that a rigorous functional
analysis, MaxEnt treatment,   yields Eqs. (\ref{eq2.8}),
(\ref{eq2.9}), (\ref{eq3.8}), and   (\ref{eq3.9}). This entails that
Eqs.  (3) and (4) of \cite{abe} cannot be correct. Neither are
correct Eqs.  (7) and  (8)of such reference. The question is the
passage from the continuous to the discrete (or viceversa)
scenarios. This is the main issue discussed in \cite{abe}.

It is well known in Measure Theory \cite{MT} that it is not possible to pass
    from a discrete probability distribution (PD) to a continuous one via a simple
         Riemann integral, as done in \cite{abe}. Moreover, a discrete PD can be cast
as an integral over a  Lebesgue-Stieltjes measure, concentrated on a
finite (or numerable) set of points. For instance, in the  Shannon
entropic case one has, in phase space ($q,p$), with $U$ the energy and $\lambda_i$ Lagrange multipliers:

\begin{equation}
\label{eq4.1}
S=-\sum\limits_{i=1}^n P_i\ln P_i=
-\int\limits_MP[U(p,q),\lambda_1.....,\lambda_n]
\ln P[U(p,q),\lambda_1.....,\lambda_n] d\mu(p,q),
\end{equation}
with
\begin{equation}
\label{eq4.2}
d\mu(p,q)=\sum\limits_{i=1}^n\delta[U(p,q)-U_i]dU(p,q),
\end{equation}
where
\begin{equation}
\label{eq4.3}
P_i=P[U_i,\lambda_1.....,\lambda_n].
\end{equation}
Thus, it becomes clear that Eq. (7) of  \cite{abe} is incorrect because there, the probability density
is NOT concentrated on a numerable set of points. The Tsallis' entropy scenario is identical to the one above. One just must insert in (\ref{eq4.1})  $P^q ln_q P$ instead of $P \ln{P}$.

\vskip 3mm
\nd Note also that Eq. (11) of \cite{abe} is NOT the standard classical Boltzmann's entropy expression, transcribed in his tombstone at Vienna's cemetery. This, of course, does not contain Planck's constant. The derivation that follows such Eq. (11) is thus invalid.

\vskip 3mm
\nd Finally, we cite some recent works that successfully deal with the Tsallis' q-statistics with continuous probability distributions \cite{one,two,three}.

\section{Conclusions}

We have here shown, by recourse to Functional Analysis, that Tsallis's thermostatistics is fully valid in the continuous instance, notwithstanding the arguments of  \cite{abe}.

 \newpage



\end{document}